\RequirePackage{fix-cm}
\documentclass[twocolumn]{svjour3}          % twocolumn
\smartqed  % flush right qed marks, e.g. at end of proof
\usepackage{graphicx}
\usepackage{amsmath}
\begin{document}

\title{Cooper pairs at above-critical current region%\thanks{Grants or other notes
}


\author{Yongle Yu        % \and
        %Second Author %etc.
}


\institute{Yongle Yu \at
              Wuhan Institute of Physics and Mathematics, 
              Chinese Academy of Sciences, Wuhan, 430071, China \\
                \email{yongle.yu@gmail.com}           %  \\
}

\date{Received: date / Accepted: date}

\maketitle

\begin{abstract}
It is generally believed that in a superconducor 
Cooper pairs are broken at above-critical current region, 
corresponding to the lost of superconductivity. 
We suggest that,  
under some circumstance, Cooper pairs
could still exist above critical current,
 and that dissipation of the system 
 is caused by the scattering of these pairs.
The existence of Cooper pairs in this region can be 
revealed by investigating the temperature dependence of 
the electrical resistance.
\end{abstract}

\section{Introduction}

The formation of Cooper pairs in a superconductor
is essential for superconductivity \cite{pair}\cite{BCS}.
Cooper pairs are of bosonic nature, and it is
believed that they generate a
supercurrent in the same mechanism as helium 
atoms generate a superflow. It is clear that,
once the cooper pairs in a superconductor are
broken, superconductivity will be lost. An intersting
question is that, can the system become dissipative
without the breaking of cooper pairs?

From the view point of many-body physics,
one can see there is a rather striking 
difference between the origin of Cooper pairs
and the origin of superconductivity.
Cooper pairs is due to some {\it attractive} 
interactions between fermions, while 
superconductivity, like superfluidity, has something
to do with {\it repulsive} interactions between 
its composing bosons, {\it i.e.}, the Cooper 
pairs \cite{sf}\cite{leggett}. This difference implies that
the physical regime of Cooper pairs is not exactly
the same as that of superconductivity.
We suggest that, under some circumstances, 
Cooper pairs are not broken when 
one increases a supercurrent to above 
critical current. We shall
also show that, at the above-critical current region,
the electrical resistance, caused by the scattering
of cooper pair,  has a different
temperature dependence from the  case where 
the charge carriers of the current are fermions 
(electrons or holes). Thus the existence of Cooper 
pairs can be signified by
the temperature behavior of the
resistance.

  \begin{figure} %[ht]
\includegraphics[width=3in]{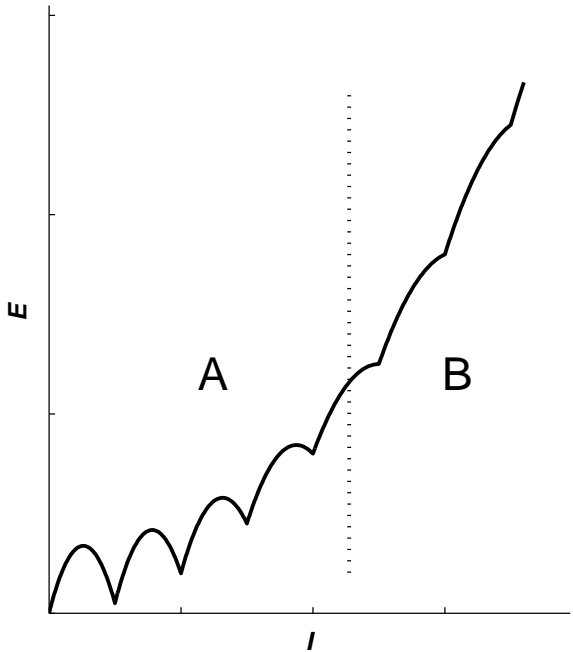}
\caption{ A schematic plot of the many-body 
dispersion spectrum of the charge carrier
 system in a superconductor.
Region $A$ is the superconducting region where 
there are metastable current-carrying states 
corresponding to the local minima
of the curve;  Region $B$ is dissipative without
 metastable states. The many-body eigen energy
states at region $A$ can be mapped to the states 
at region $B$ by Galileo
transformation.  }
\label{dispersion}
\end{figure}
Like the case of superfluidity 
\cite{sf}\cite{leggett}\cite{bloch}, the dissipative behavior of
a superconductor above a critical current $I_c$,
can be explained naturally in terms of   
the many-body dispersion spectrum $E(I)$ 
of the charge carrier
system  ($E(I)$ is the
 lowest eigen energy at given current $I$). Beyond $I_c$, 
$E(I)$ is a monotonically 
increasing function of $I$ ( see Fig.~\ref{dispersion}), 
thus  a current can continuously lose its
energy and momentum to the environment, 
corresponding to a dissipative decay process. 
This is contrast to the case 
in the $I<I_c$ regime where the supercurrents
are metastable states, corresponding to the local minima
of the $E(I)$ curve, whose decay is prevented by the 
energy barriers among the minima.

The dissipation mechanism illustrated above  
does not involves breaking of Cooper pairs. 
Moreover, the low-lying  eigen energy states below
$I_c$ can be Galileo boosted to generate
 the low-lying
 states above $I_c$ \cite{sf}\cite{bloch} 
 \cite{galileo}. The Galileo boost, 
 also could being viewed as 
 center-of-mass-mention  transformation, 
 does not modify the inner structure
 of the states such as pair correlations. 
 Thus, the Cooper pairs,  present in states
 below $I_c$, survive the
 boost and  exist in the states 
 above $I_c$ \cite{mix}.
  
% the states below critical current. Let the many-body
% eigenstates below critical current is denoted by
%$\psi_{I<I_c}(\boldsymbol{r}_1,\sigma_1,
%\boldsymbol{r}_2,\sigma_2, ...)$ where 
%$\boldsymbol{r}_i,\sigma_i$ is the coordinates and
%the spin of $i^{th}$ electrons, respectively. Then
%$\Psi (\boldsymbol{r}_1,\sigma_1,
%\boldsymbol{r}_2,\sigma_2, ...) = e^ {i \boldsymbol{k}_1 \dot
%\boldsymbol{r}_1  + i\boldsymbol{k}_2\dot \boldsymbol{r}_2 + ...} 
%\psi_{I<I_c}(\boldsymbol{r}_1,\sigma_1,
%\boldsymbol{r}_2,\sigma_2, ...) $ 
%is another eigenstate
%with a momentum being the momentum of $\psi_{I<I_c}$ 
%plus $N\hbar \boldsymbol{k}$. 

In literature, Landau's criterion
of a superconductor is generally used for the analysis
of transitional regime, which
determines a critical velocity to be $ \Delta /p_F$, where  
$\Delta$ is the pairing gap and $p_F$ is the Fermi momentum.
Landau's criterion requires an exchange of a quantum, 
with a large momentum (the order of $p_F$)
and with a certain energy,   between the superconducting
charges and its environment. However, this exchange 
process could be inhibited for that the spectrum of 
the environment may generally be incompatible to absorb
such an unusual quantum.

\begin{figure} %[ht]

\includegraphics[width=3in]{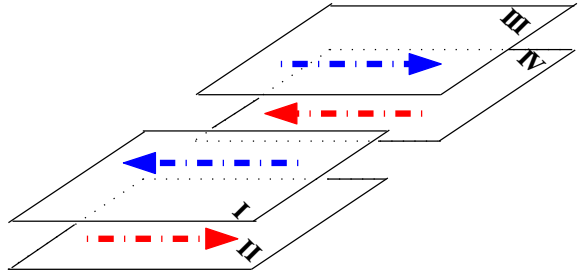}

\caption{\label{currentalignment} To weaken the Silsbee effect,
four thin superconducting films are used to carry the current.
 Film I is on top of film II, and film III on top of film IV. Film I
 is in the same plane of film III, and film II in the same plane of film
 IV.  Current in the film I (blue arrow) points to 
 the same direction as to the current
 in film IV (red arrow), while the currents in the film II (red arrow) 
 and film IV (blue arrow) point to the opposite
 direction. }
\end{figure}

In a superconductor, the self-induced magnetic field of
the supercurrent could lead to loss of 
 superconductivity by breaking cooper pairs, ({\it i.e.},  
 Silsbee effect \cite{silsbee}).  In this case, there is
 an 'external' critical current determined by the critical
 magentic field, and the Cooper pairs do not exist
 above critical current. However, Silsbee 
 effect can be largely reduced by alignment of the currents 
 and their geometries (see Fig.~\ref{currentalignment} ), 
 thus one might be able to obtain an
 intrinsic critical supercurrent, like the case 
 of superfluid $^4$He,   
 and reach a dissipative regime without breaking Cooper pairs.  
 We shall assume
 the existence of Cooper pairs  above critical current  
 in follows.
 
\section{Analysis and Results}

The eletrical resistance in a superconductor above $I_c$
and below $T_c$ is mainly caused by the scattering properties of 
 Cooper pairs with phonons. We shall discuss the temperature
 behavior of the resistance. 
For simplicity, we assume 
physical properties of the superconductor are
isotropic. First, we consider the
dispersion relation of a Cooper pair.
The dispersion is linear at
 small momentum $\boldsymbol{q} $, {\it i.e.},  
$\varepsilon(\boldsymbol{q}) \approx v |q|$ ( we 
approximate $v$  by the critical
 velocity of the supercurrent $v_c$).
At large $\boldsymbol{q}$, the 
energy of a Cooper pair is approximately 
$q^2/2m_c$ where $m_c$ is the mass of a Cooper 
pair assumed to be $2m_e$ ($m_e$ is the mass of an electron). 
A general form of dispersion
\begin{equation}
\varepsilon (\boldsymbol{q})= \sqrt{v_c^2 q^2 + (q^2/4m_e)^2}
\end{equation}
can be used for approximation for all $\boldsymbol{q}$
values \cite{wen}.

\begin{figure} %[ht]
\includegraphics[width=3in]{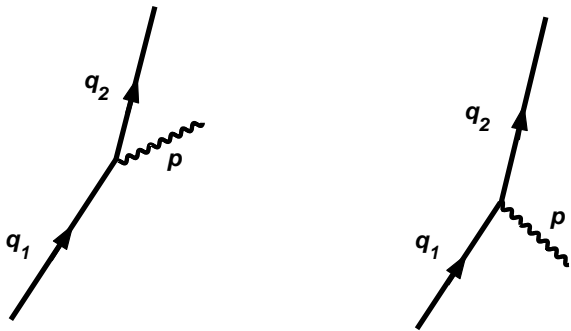}
\caption{One phonon scattering process
of a Cooper pair. left (right) corresponds
to the emission (absorption) of a phonon.  }
\label{onePhonon}
\end{figure}

We consider the leading scattering process in which one phonon 
is adsorbed or emitted by a Cooper pair (see Fig \ref{onePhonon}). 
The total energy and momentum should be conserved,
\begin{equation}
\varepsilon (\boldsymbol{q_1})= 
  \varepsilon(\boldsymbol{q_2}) \pm \hbar c_s p 
\end{equation}

\begin{equation}
\boldsymbol{q_1}= \boldsymbol{q_2} \pm \boldsymbol{p}
\end{equation}
Where $\boldsymbol{q_1}$ ($\boldsymbol{q_2}$) is the momentum
of a cooper pair before (after) the scattering, $\boldsymbol{p}$
is the momentum of the phonon, $c_s$ is the sound velocity,
and $+$ ($-$) corresponds to the emission (absorption) of 
the phonon.

Combining Eq. 1, Eq. 2 and Eq. 3, one gets,
\begin{equation}
\sqrt{v_c^2 q_1^2 + (q_1^2/4 m_e)^2} = 
\sqrt{v_c^2 q_1^2 + (q_1^2/4 m_e)^2} \pm \hbar c_s 
|\boldsymbol{q_1}- \boldsymbol{q_2}| 
\end{equation}

 In most superconductors,  $v_c$,  can being estimated 
 using critical current density \cite{vs_current},
 is roughly the orders of ten meters per second or below.  
 Thus $v_c$ is orders
of magnitude smaller than $c_s$. 
 One then can realize that
unless the $q_1$ ($q_2$) in the phonon emission (absorption) process  
is roughly equal
 or larger than $2 m_e c_s$, Eq. 4 can not be satisfied.
 for a Cooper pair with a momentum of $ 2 m_e c_s$, the 
 energy is roughly $ m_e c_s^2$ and the corresponding
 temperature is $T^* =   m_e c_s^2/ k_B$ where
 $k_B$ is the Boltzmann constant. Thus at low temperature
 $T< T^*$, one-phonon scattering process is absent, and
 the scattering process of Cooper pairs must involve at 
 least two phonons.
 
 One can invoke Boltzmann transport equation to determine
 the temperature dependence of resistance of the Cooper pair
 system. However, unlike the case of a fermionic system where
 the existence of Fermi surface and Pauli blocking are
 essential, the analysis of the (bosonic) Cooper pair system can
 be simplified. At $T>T^*$, the one-phonon scattering process
  can bring the momentum of a Cooper pair
 (with an energy of $k_B T$ or less)  to zero or to the 
 opposite direction,  thus effectively causing the
 dissipation  of the current. The probability of
 one phonon process in low temperature is proportional to
 $T$ (see, {\it e.g.}, \cite{mott}), thus the resistance
 $R$ is linear in $T$. At $T<T^*$, one can find that
  probability of two-phonons process is proportional to
 $T^2$, thus $R \propto T^2$.  In metals where
 temperature dependence of resistance is
  determined by electron phonon scattering,
 the power law of $R(T)$ at low temperature 
 is different, with an exponent being 5 
 (see, {\it e.g.}, \cite{mott}\cite{lead5}\cite{lead5_2}). Thus one 
 can distinguish between Cooper pairs and electrons
 by checking the powering law
 of $R(T)$ in the above critical current regime.

In the so called strange metal phase of some high-$T_c$ cuprates,
the resistance has also a linear temperature dependence. One might
wildly speculate that this linear behavior could be
caused by the scattering of Cooper pairs. In order for 
this speculation
 to be valid, Cooper pairs shall exist 
above $T_c$ in some systems. An example is that cold Fermi
 atom gases which
can be tuned to go through BCS-BEC crossover. In the BEC side,
the binding energy of  Cooper pairs (or molecules) can be
orders of magnitude larger than 
superfluid transition temperature $T_c$,
since $T_c$ can be made small by decreasing 
the density of atom gases.

\section{Conclusions}
we suggest that, under some circumstance, 
Cooper pairs could exist in
a superconductor above critical current, 
and the system has a
different power law of $R(T)$ from the
case where electrons are current carriers.

\begin{acknowledgements}
This work was supported by Chinese NSF 
Grant No. 11075201.
\end{acknowledgements}


\begin{thebibliography}{}
\bibitem{galileo} In superconductors, Galileo invariance is not
exact but is a good approximation when the center of mass motion
is not large.
\bibitem{mix} there might be some possibility that the dispersion
 energy of normal  state (without pairing) could be smaller than
 the dispersion energy of pairing state within  a
 current regime somewhere between $I_c=0$ and the first supercurrent, 
 we ignore this case here.
\bibitem{vs_current}  We obtain $v_c$ using $j_c = n_c * (2e)* v_c$,
 where $j_c$ is the critical current density,  $n_c$ is
 the density of Cooper pairs, $2e$ is the charge of a Cooper pair.
 We assume that all conducting electrons are paired up, thus
 $n_c = n_e/2$,  where $n_e$ is the density of electrons.
\end{thebibliography}


\end{document}